\documentclass[aps,prl,twocolumn,superscriptaddress]{revtex4}
\usepackage{graphicx}

\begin{document}


\title{Magnetic and superconducting properties of Cd$_2$Re$_2$O$_7$: Cd NMR and Re NQR.}

\author{O.~Vyaselev}
\affiliation{Institute for Solid State Physics, University of Tokyo 5--1--5
Kashiwanoha, Kashiwa, Chiba 277--8581, Japan}

\affiliation{Institute of Solid State Physics Rus. Ac. Sci., Chernogolovka Mosc. Distr., 142432 Russia}
\author{K.~Arai}\author{K.~Kobayashi}\author{J.~Yamazaki}\author{K.~Kodama}
\author{M.~Takigawa}
\author{M.~Hanawa}\author{Z.~Hiroi}
\affiliation{Institute for Solid State Physics, University of Tokyo 5--1--5 Kashiwanoha, Kashiwa, Chiba 277--8581,
Japan}

\date{\today}

\begin{abstract}
We report Cd NMR and Re NQR studies on Cd$_2$Re$_2$O$_7$, the first superconductor among pyrochlore oxides ($T_c
\simeq \mathrm{1~K}$). Re NQR spectrum at zero magnetic field below 100~K rules out any magnetic or charge order.
The spin-lattice relaxation rate below $T_c$ exhibits a pronounced coherence peak and behaves within the
weak--coupling BCS theory with nearly isotropic energy gap. Cd NMR results point to moderate ferromagnetic
enhancement at high temperatures followed by rapid decrease of the density of states below the structural
transition temperature of 200~K.
\end{abstract}

 \maketitle

The pyrochlore transition metal oxides with the chemical formula A$_2$B$_2$O$_7$ known for decades have recently
attracted renewed interest.  A and B sublattices both form a network of corner--sharing tetrahedra.  The geometry
of such a lattice causes strong frustration of magnetic interaction when occupied by local moments, resulting in
large degeneracy of the low energy states and anomalous magnetic behavior \cite{Ramirez991}.  Interplay between
strong correlation and geometrical frustration in itinerant electron systems is a challenging issue.  For example,
large degeneracy due to frustration is considered to be crucial also for the heavy--electron behavior of the
spinel compound LiV$_2$O$_4$ \cite{Urano001}, where V sites form the identical pyrochlore lattice.  Recent
investigations of pyrochlores including 5$d$ transition metal elements have led to the discovery of the
superconductivity, for the first time among pyrochlore oxides, in Cd$_2$Re$_2$O$_7$ below $T_c \simeq
\mathrm{1~K}$ \cite{HiroiSC, SakaiSC,MandrSC} and unusual metal--insulator transition in Cd$_2$Os$_2$O$_7$
\cite{Mandr01}.

Besides superconductivity, Cd$_2$Re$_2$O$_7$ shows an unusual phase transition at 200~K \cite{HiroiSC,HiroiStr,
MandrStr}.  Although complete structural determination below 200~K is yet to be done, X--ray studies proposed a
second--order transition from the high--temperature (high--$T$) cubic $Fd\bar{3}m$ space group to the low--$T$
cubic $F\bar{4}3m$ space group \cite{HiroiStr}, accompanied by loss of inversion symmetry at Re and Cd sites. The
transition also causes a large change of electronic properties \cite{HiroiSC, MandrStr}.  The resistivity shows
almost no $T$--dependence near room temperature but drops abruptly below 200~K. The magnetic susceptibility also
decreases steeply below 200~K, while at higher temperatures it shows only weak $T$--dependence with a broad
maximum near 300~K. Moreover, a kink in resistivity near 120~K indicates an additional phase transition
\cite{HiroiStr, HiroiStr2}. Results of band structure calculations \cite{Harima011,SinghCM} indicate that the
material is a compensated semi-metal with  two electron (one hole) Fermi surfaces centered centered at the 
$\Gamma$ (K) point of the Re-5$d$($t_{2g}$) bands.  
The electronic structure near the Fermi level is sensitive to structural distortion.  While it is proposed that a
structural transition may occur to remove large spin degeneracy of localized moments on a pyrochlore lattice
\cite{Yamashita001}, relation between structure and electronic properties is largely unexplored for itinerant
electron systems.  

In this letter we describe the microscopic information on the magnetic and superconducting
properties of Cd$_2$Re$_2$O$_7$ obtained from the nuclear magnetic resonance (NMR) at Cd sites and the nuclear
quadrupole resonance (NQR) at Re sites. The procedure of growing single crystals of Cd$_2$Re$_2$O$_7$ has 
been described earlier \cite{HiroiSC}. NMR measurements were done on a single crystal in the external field of 
10.47~T. For NQR, a crystal was crushed into powder. Standard spin--echo pulse techniques were used, including 
inversion--recovery to measure spin--lattice relaxation rate. 

\begin{figure}[b]
\includegraphics[scale=0.35]{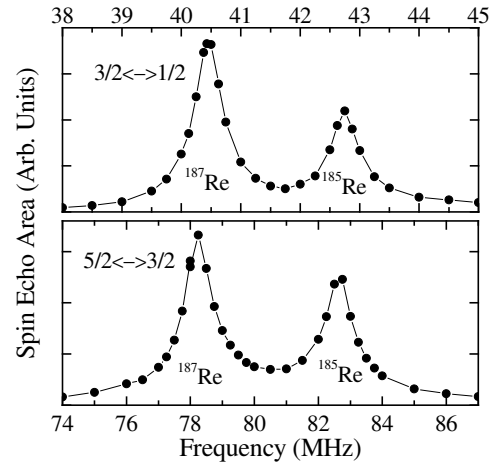}
\caption{\label{ReNQR}Re NQR spectra at 5~K. Top panel: $\pm3/2\leftrightarrow\pm1/2$ transition. Bottom panel:
$\pm5/2\leftrightarrow\pm3/2$ transition.}
\end{figure}
We first discuss the low-temperature NQR results on Re sites.  NQR spectra of $^{185}$Re and $^{187}$Re nuclei,
both with spin 5/2, have been obtained below 100 K. Measurements at higher temperatures were hampered by short spin--spin
relaxation time. The spectra at 5~K are shown in Fig.~\ref{ReNQR} where the top and the bottom panels correspond
to $\pm3/2\leftrightarrow\pm1/2$ and $\pm5/2\leftrightarrow\pm3/2$ transitions, respectively. NQR frequency of
spin 5/2 nucleus can be expressed as \cite{Abrag}: $\nu_{\pm3/2,\pm1/2}\approx\nu_{Q}[1+(59/54)\eta^{2}]$ and
$\nu_{\pm5/2,\pm3/2}\approx2\nu_Q[1-(11/54)\eta^{2}]$, where $\nu_{Q} \propto V_{zz} = \partial ^2 V/\partial z
^2$ is the largest principal value of the electric field gradient (EFG) tensor at the nuclear position, $V$ is the
electrostatic potential, and $\eta=|V_{xx}-V_{yy}|/|V_{zz}|$ is the asymmetry parameter. Applying these expressions
to the measured spectra one finds $\nu_{Q}=40.4$~MHz at 5~K, gradually decreasing with temperature to 37.6~MHz at
100~K, and $\eta=0.166\pm0.002$ without noticeable temperature dependence.

A sharp line for each NQR transition of both Re isotopes immediately rule out non--uniform charge distribution
among Re sites such as charge density wave states.  Because of the large quadrupole moments of $^{185,187}$Re and
high sensitivity of the EFG to the local charge distribution of 5$d$ electrons, any non--uniform charge
distribution should give rise to broadening or splitting of the NQR lines.  We can also rule out any magnetic
order, since 1~$\mu_B$ of spin moment in $5\it{d}$ state typically gives the hyperfine field of 100~T.  Thus even
a tiny moment would have caused the Zeeman splitting of the NQR spectra.

Re sites possess three--fold rotation symmetry both in the high--$T$ $Fd\bar{3}m$ structure or in the $F\bar{4}3m$
structure, the latter being proposed by X--ray for the structure below 200~K \cite{HiroiStr}.  Since EFG is a
second rank symmetric tensor, it must be axially symmetric in these cases.  Therefore, the non--zero value of $\eta$
implies further lowering of symmetry at least in the temperature range below 100~K.

\begin{figure}[t]
\includegraphics[scale=0.3]{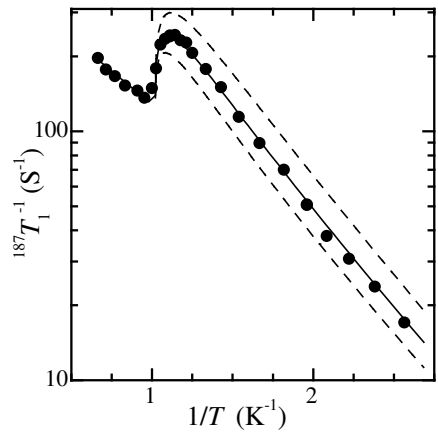}%
\caption{\label{BCST1}  $^{187}$Re spin--lattice relaxation rate {\it vs} inverse temperature.  The
solid line is the fit to the weak coupling BCS theory with $\delta/\Delta$=0.22, $\Delta(T=0)$=1.80~K, and
$T_c$=0.98~K. The dashed lines represent the cases with $\delta/\Delta$=0.27 or 0.17}
\end{figure}

Fig.~\ref{BCST1} shows the spin--lattice relaxation rate $^{187}T_1^{-1}$ of $^{187}$Re nuclei measured on the
$\pm5/2\leftrightarrow \pm 3/2$ NQR transition at zero magnetic field as a function of inverse temperature. The
relaxation was confirmed to be of magnetic (not quadrupolar) origin at several temperatures by observing that the
ratio of $T_1^{-1}$ for the two Re isotopes is equal to the squared ratio of the nuclear magnetic moments.

Above $T_c$ in the region 1~--~5~K, $^{187}T_{1}^{-1}$ is practically linear in temperature,
$^{187}(T_1T)^{-1}$~=~127 (Sec$\cdot$K)$^{-1}$.  Just below $T_{c}$, $^{187}T_{1}^{-1}$ increases sharply
exhibiting a strong coherence peak \cite{HeSl} with a maximum of about 245~(Sec$\cdot$K)$^{-1}$ at $\sim0.88$~K,
which is twice as large as just above $T_{c}$.  Below 0.8~K the relaxation rate decreases following an activated
$T$--dependence.

It is well known that a coherence peak in the $T$--dependence of $T_1^{-1}$ is easily depressed when magnitude of
the superconducting gap varies substantially or the order parameter changes sign over the Fermi surface. Thus the
observed well--pronounced coherence peak provides a clear evidence for nearly isotropic $s$--wave superconducting gap.
Nodeless gap in this material have been recently inferred also from specific heat
data \cite{HiroiSC2} and $\mu$SR \cite{MandrMuSR, Kadono} measurements of the penetration depth. We have fitted
the data to the BCS expression, considering distribution of the gap due to possible variation over the Fermi surface
\cite{MacLaughlin}. We assumed a uniform distribution of the gap between $\Delta - \delta$ and $\Delta + \delta$ and
$\delta/\Delta$ to be
independent of temperature.  The $T$-dependence of $\Delta$ is assumed to follow the weak coupling BCS theory.
A good fit was obtained for $\delta/\Delta$=0.22, $\Delta(T=0)$=1.80~K, and $T_c$=0.98~K as
shown by the solid line in Fig.~\ref{BCST1}. The ratio $\Delta(T=0)/T_c$=1.84 is close to the weak coupling
BCS value 1.75, justifying our analysis. The dashed lines in Fig.~\ref{BCST1} indicates that the height of the
coherence peak is sensitive to distribution of the gap.  The fitted value of $\delta/\Delta$ puts the upper limit for the gap variation,
since any damping of the quasi-particles will also suppress the coherence peak.

Let us now turn to the NMR results on Cd sites.  The Knight shift, $K$, and the spin--lattice relaxation rate,
$^{111}T_1^{-1}$ were measured for $^{111}$Cd nuclei (spin 1/2) using a single crystal. In the high temperature
$Fd\bar{3}m$ structure, three--fold rotation symmetry at Cd sites should give rise to the axially symmetric Knight
shift, $K=K_{iso}+K_{ax}(1-3\cos^2\theta)$, where $\theta$ is angle between the external field and the [111]
direction.  This angular dependence was observed above 200~K. Upon cooling below the structural transition
temperature (200~K) the line splits in two. The average shift of the split lines follows the same angular
dependence and thus the values of the isotropic ($K_{iso}$) and the axial ($K_{ax}$) parts of the shift were
determined as a function of temperature. Since the splitting is very small (less than 0.018\%), the conclusions of
the following analysis do not depend on the validity of such averaging.
\begin{figure}[t]
\includegraphics[scale=0.35]{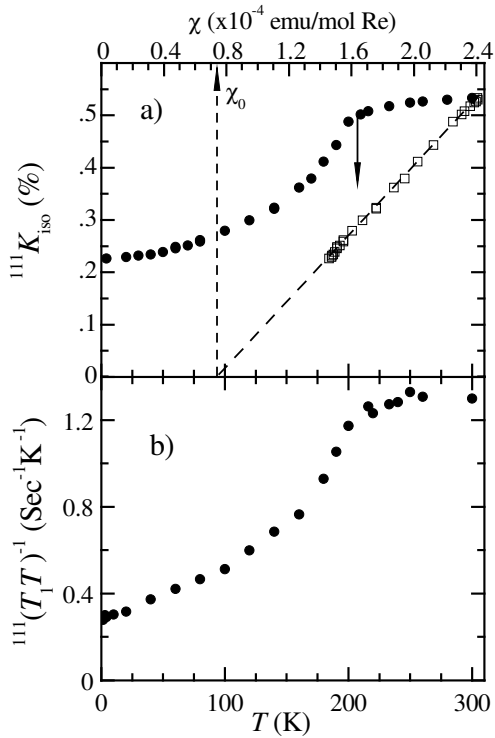}%
\caption{\label{CdNMR} (a): Isotropic part of $^{111}$Cd Knight shift {\it vs} temperature (circles -- bottom axis),
and {\it vs} magnetic susceptibility (squares -- top axis). (b): Temperature dependence of $(T_1T)^{-1}$ at
 $^{111}$Cd sites.}
\end{figure}

In Fig.~\ref{CdNMR}a, $K_{iso}$ is plotted against temperature (bottom axis) or magnetic susceptibility, $\chi$
(top axis). The axial part $K_{ax}$ (not shown in the figure) is --0.042\% at 300~K, much smaller than $K_{iso}$,
and decreases only by 0.01\% upon cooling down to 4~K, indicating that the dominant hyperfine field comes from the
Cd $s$--states hybridized with the Re 5$d$ conduction states. Similar to the magnetic susceptibility $\chi$
\cite{HiroiSC}, $K_{iso}$ decreases rapidly below 200~K. The shift is generally expressed as
$K_{iso}=K_0+K_{s}(T)$, where the first term is the $T$--independent chemical shift and the second term is the
$T$--dependent spin shift from conduction electrons. Likewise $\chi=\chi_0+\chi_{s}(T)$, where the first term is
the sum of the diamagnetic and the orbital (Van Vleck) susceptibility. As shown in Fig.~\ref{CdNMR}, linear
relation was found between $K_{iso}$ and $\chi$ in the whole temperature range, yielding the hyperfine coupling
constant $A_{hf}=N_A\mu_{B}K_{s}/\chi_{s}=17.9$~T$/\mu_{B}$. Since the chemical shift of Cd in a non--magnetic and
non--metallic substance is typically 0.01~\% or less \cite{CdKs} and much smaller than the observed value of
$K_{iso}$, we assume $K_0$=0 in the following analysis.  We then obtain $\chi_0=0.75\times10^{-4}$~emu/mole~Re,
leading to $\chi_{s}=1.65\times 10^{-4}$~emu/mole~Re at 300~K. From the standard value for the diamagnetic
susceptibility $\chi_{dia}=-0.76\times 10^{-4}$ emu/mole~Re, the orbital susceptibility is estimated as
$\chi_{orb}=1.51\times 10^{-4}$ emu/mole~Re.

In Fig.~\ref{CdNMR}b,  is shown the $T$--dependence of $^{111}(T_1T)^{-1}$ measured for $H\| [001]$. At some
temperatures $^{111}T_1$ was measured for $H\| [110]$ and found to be isotropic. We found that $^{111}(T_1T)^{-1}$
also shows rapid reduction below 200~K, indicating that the phase transition at 200~K causes sudden loss of the
density of states (DOS). For non--interacting electrons, when the spin shift and the spin--lattice relaxation are
both due to contact interactrion with unpaired $s$--electrons, the {\it Korringa relation} is known to be satisfied,
$T_1TK_{s}^2 =(\gamma_{e}/\gamma_{n})^2(\hbar/4 \pi k_B) \equiv S$, where $\gamma_{e}$ and $\gamma_{n}$ are
respectively electronic and nuclear gyromagnetic ratios.  In real metals, the ratio\begin{equation}
K_{\alpha}=S/(T_{1}TK_{s}^{2})  \label{eqnT1TK2}
\end{equation}
provides a useful measure of magnetic correlation \cite{MoriyaRPA, NarathRPA}.  Since $(T_1T)^{-1}$ probes the
dynamical susceptibility averaged over the Brillouin zone, it can be enhanced by either ferromagnetic (FM) or
antiferromagnetic (AF) spin correlations, while only FM correlation strongly enhances the spin shift. Thus,
generally speaking, the value of $K_{\alpha}$ being much smaller (larger) than unity is a signature for
substantial FM (AF) correlation.

\begin{figure}[t]
\includegraphics[scale=0.3]{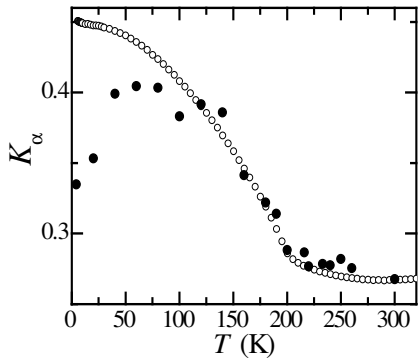}
\caption{\label{Kal} Temperature dependence of $K_{\alpha}$. Filled circles: experimental data of $K_{\alpha}$
obtained from $^{111}$Cd $(T_1T)^{-1}$ and $K_{spin}$=$K_{iso}$ using eqn.~(\ref{eqnT1TK2}) Open circles:
$K_{\alpha, RPA}$ calculated from eqs.~(\ref{eqnchis}, \ref{eqnKa}) and the data of $\chi_{s}$ from
ref.~\cite{HiroiSC}.}
\end{figure}

Solid circles in Fig.~\ref{Kal} show the $T$--dependence of $K_{\alpha}$.  The value of $K_{\alpha}\sim 0.28$
above 200~K points to moderate FM enhancement. The sudden increase of $K_{\alpha}$ below 200~K can be accounted
for by loss of DOS as explained below.  However, there is a broad peak in $K_{\alpha}$ near 60~K,  and at lower
temperatures $K_{\alpha}$ decreases.  Choice of non-zero values for $K_0$ does not change this behavior, which is
already evident in Fig.~\ref{CdNMR} since $^{111}(T_1T)^{-1}$ keeps decreasing whereas $K(T)$ becomes nearly flat
below 50~K.

In order to examine to what extent the behavior of $K_{\alpha}$ is explained by $T$--dependence of DOS, we utilize a
simple RPA expression with a free--electron--like dispersion of the conduction band.
This analysis is mainly for qualitative purpose since specific feature of the multiple Fermi surfaces
of Cd$_2$Re$_2$O$_7$ \cite{Harima011, SinghCM} is not considered. We also neglect possible variation of the matrix elements 
of the contact interaction over the Fermi surface. Although the good proportionality between $K$ and $\chi$ supports such a 
simple situation, careful examination would be required in future in the light of realistic Fermi surface and wave functions.
The Stoner enhancement factor of the spin susceptibility
\begin{equation}
1/(1-\alpha)=\chi_{s}/(2\mu_{B}^2\rho_0) , \ \  \alpha=U\rho_0 , \label{eqnchis}
\end{equation}
where $\rho_0$ is the DOS at the Fermi level and $U$ denotes the Coulomb repulsion, is related to $K_{\alpha}$ by RPA as
\cite{MoriyaRPA,NarathRPA}
\begin{equation}
K_{\alpha, RPA}=2\int_0^1 \frac{(1-\alpha)^2xdx}{[1-\alpha G(x)]^2} ,\label{eqnKa}
\end{equation}
$G(x)=0.5\{1+(1-x^2)/(2x)\}\ln |(1+x)/(1-x)|$. Applying eq.~(\ref{eqnKa}) to the experimental data of
$K_{\alpha}$, we obtain $1/(1-\alpha)=7.3$ at 300~K.  This value is comparable to the enhancement factor of 4.4 obtained from
the calculated DOS value for the room temperature structure (0.58~eV$^{-1}$ per Re \cite{Harima011}) and the experimental
value of $\chi_s$ obtained above.

We now determine the $T$--dependence of DOS or $\alpha$ from the experimental data of $\chi_{s}(T)$ by using
eq.~(\ref{eqnchis}), $\chi_{s} \propto \alpha/(1-\alpha)$, with the initial condition that  $1/(1-\alpha)=7.3$ at
$T$=300~K.  We then calculate $K_{\alpha, RPA}$ using eq.~(\ref{eqnKa}). The results are shown by the circles in
Fig.~\ref{Kal}.  One can see that from high temperatures down to $\sim$120~K  the calculated $K_{\alpha, RPA}$
reproduces the experimental data fairly well, supporting our assumption that loss of DOS is the major cause for
the reduction of both $\chi_{s}$ and $^{111}(T_1T)^{-1}$ below 200~K. However, they diverge below 120K.
The anomalous behavior at low temperatures implies that either ferromagnetic enhancement is restored or there
is additional suppression only for the dynamical susceptibility.

We found that the ratio of $T_1^{-1}$ at Re and Cd sites is unexpectedly large. Hyperfine
interaction between Re nuclei and the 5$d$ conduction electrons should be mainly due to core polarization effect,
with typical magnitude --120T/$\mu_B$ \cite{Shaham}. From the values of the hyperfine coupling, the nuclear
gyromagnetic ratio and the reduction factor for the relaxation due to core polarization field \cite{Yafet}, we expect the
ratio $^{187}T_1^{-1}/^{111}T_1^{-1}$ to be 17. Experimentally we obtain
$^{187}T_1^{-1}/^{111}T_1^{-1}$=420 at 5K. Possible reasons for this may be the following:  (1) Orbital hyperfine
field gives large contribution to $^{187}T_1^{-1}$. (2) The hyperfine coupling, in particular for Cd nuclei, is
not uniform over the Fermi surface, i.e. some part of the Fermi surface do not couple effectively to Cd nuclei but
makes large contribution to the Re relaxation rate.  The latter case points to non--trivial magnetic correlation
among 5$d$ electrons which does not manifest in the Cd relaxation data.

Finally, we comment on the large reported value of the $T$-linear coefficient of the specific heat
$\gamma$=15~mJ/K$^2$ above $T_c$ \cite{HiroiSC}.  Our estimate of $\chi_0$ and the susceptibility data in
ref.~\cite{HiroiSC} give $\chi_{s}=7\times 10^{-5}$~emu/mole~Re at 5~K.  We then obtain the Wilson ratio
$R_W$=0.34. Such a small value of $R_W$ would normally imply strong electron--phonon coupling and appears to be
incompatible with the weak coupling superconductivity.  A exotic possibility may be that the large $\gamma$ is due
to low lying excitations which do not contribute to $\chi_s$, for example, orbital fluctuations.

In summary, Re NQR data below 100 K reveals no sign of magnetic or charge order in Cd$_2$Re$_2$O$_7$.
However, asymmetry of electric field gradient indicates lack of three--fold rotational symmetry. The superconducting
state according to the temperature dependence of Re NQR spin--lattice relaxation rate, is apparently a
weak--coupling BCS case with a nearly isotropic energy gap. Cd NMR data show moderate ferromagnetic
enhancement above 200K.  Rapid decrease of the spin susceptibility and $^{111}(T_1T)^{-1}$ below the structural
phase transition at 200~K is accounted for by loss of the density of states (DOS) down to 120K.  The shift and relaxation
rate data at lower temperatures pose puzzles which remain to be clarified.

\begin{acknowledgments}
We would like to thank H.~Harima and K.~Ueda for enlightening discussion. This work is supported by the Grant--in--Aid for
Scientific Research, No.~12--00037 and  No.~10304027 from the Japan Society for the Promotion of Science (JSPS), and
No.~12046219, Priority Area on {\it Novel Quantum Phenomena in Transition Metal Oxides}) from the Ministry of Education,
Culture, Sports, Science and Technology Japan. The research activity of O.V. in Japan is supported by the
JSPS Postdoctoral Fellowship for Foreign Researchers in Japan.
\end{acknowledgments}

%


\end{document}